\documentclass[aps,prb,preprint,showpacs,preprintnumbers,amsmath,amssymb]{revtex4}
\usepackage{graphicx}
\usepackage{float}
\begin{document}
\title{Role Played by Surface Plasmons on Plasma Instability in Composite Layered Structures}

\author{Godfrey Gumbs$^{1,2}$, Andrii Iurov$^{1}$\footnote{E-mail contact: \texttt{theorist.physics@gmail.com}} and Danhong Huang$^3$}
\affiliation{$^{1}$Department of Physics and Astronomy, Hunter College of the
City University of New York, 695 Park Avenue, New York, NY 10065, USA\\
$^{2}$ Donostia International Physics Center (DIPC),
P de Manuel Lardizabal, 4, 20018 San Sebastian, Basque Country, Spain\\
$^{3}$Air Force Research Laboratory, Space Vehicles Directorate, Kirtland Air Force Base, NM 87117, USA}

\date{\today}

\begin{abstract}
We demonstrate the engineering of a source of radiation from  growing
surface plasmons (charge density oscillations) in a composite nano-system.
The considered hybrid nano-structure consists
of a thick layer of a conducting substrate on whose surface a plasmon mode
is activated conjoining  a single or pair of thin sheets of either monolayer graphene,
silicene or a two-dimensional electron gas as would occur at a hetero-interface.
When an electric current is passed through either a layer or within the substrate,  the
low-frequency plasmons in the layer may bifurcate into separate streams due to
the driving current. At a critical wave vector, determined by the separation between
layers (if there are two) and their distance from the surface, their phase velocities
may be in opposite  directions and a surface plasmon instability leads to the emission
of radiation (spiler). Spiler takes advantage of the flexibility of choosing its
constituents to produce sources of radiation. The role of the substrate is to
screen the Coulomb interaction between two layers or between a layer and the surface.
The range of wave vectors where the instability is achieved may be adjusted by
varying layer separation and type of material. Applications to detectors and
other electromagnetic devices exploiting nano-plasmonics are discussed.
\end{abstract}
\vskip 0.2in
\maketitle

\section{Introduction}
\label{sec1}

Possible sources of terahertz (THz) radiation have been investigated for
several years now. These frequencies cover the electromagnetic spectrum
lying between microwave and infrared.
By epitaxially growing layers of different semiconductors,
high power THz quantum well  layers emitting across a wide frequency range
have been produced. The work reported so far covers ultra-long wavelength emission,
phase/mode-locking, multiple color generation, photonic crystal structures, and
improved laser performance with respect to both maximum operating temperature
and peak output power. It was  predicted by Kempa, et al.\,\cite{Bakshi} (see also
Ref.\,[\onlinecite{AI1}]) that when a current is passed through a stationary electron gas, the
Doppler shift in response frequency of this two-component
plasma leads  to a spontaneous generation of plasmon excitations
and subsequent Cherenkov radiation\,\cite{AI4star} at sufficiently high draft velocities.
Similar conclusions are expected for monolayer graphene which is
characterized by massless Dirac fermions where the energy dispersion is linear in the
wave vector ${\bf k}$ or a nanosheet of silicene consisting of silicon atoms,
which has been synthesized\,\cite{silicene}. In the same group of the periodic table
with graphene, silicene is predicted to exhibit similar
electronic properties. Additionally, it has the advantage over graphene in
its compatibility with Si-based device technologies.
The electrons in graphene with classical mobility
estimated from $\rho_{xx}=1/ne\mu_c$ (with $n$ as electron density) to be $\mu_c\sim \times 10^5\ cm^2/V\cdot s$ may
moving ballistically over distances up to   $0.2\,\mu m$.
\medskip

Plasmon modes of quantum-well transistor structures with frequencies
in the THz range may be excited with the use of far-infrared (FIR) radiation
A split grating-gate design  has been found to significantly enhance FIR
response\,\cite{2,3,gg1,gg2}. Additionally, the role played by plasma excitations in the THz
response of low-dimensional microstructures
has received considerable attention\,\cite{4,5,6,7,8,9,10,11,12,13}.
This paper discusses plasma instabilities in a pair
of Coulomb coupled layers when the  layers are  either graphene, a
two-dimensional electron gas (2DEG) or some other type of 2D-material layer in which
inter-layer hopping between layers is not included.\,\cite{12,13}
\medskip

We consider a composite nano-system
consisting of a thick layer of conducting substrate on whose surface the plasmon
is activated in proximity with a pair of thin sheets. We demonstrate how the
screening of the Coulomb coupling of the plasmons in this pair of layers
by the charge density fluctuations on the surface of a semi-infinite substrate
affects the surface plasmon instability that leads to the emission of radiation (spiler).
The excitation of these plasmon modes is induced by resonant external optical fields.
As an emitter, the spiler may be activated  optically.
Spiler exploits the
flexibility of choosing its constituents to produce coherent sources
of radiation. Applications to sensors that electromagnetic devices
exploiting nano-plasmonics are explored.
\medskip

Our approach models an ensemble consisting of a pair of 2D layers
and a thick layer of a conducting medium that emits radiation when an
electric field  splits the plasmon spectrum  which results in an
instability  when the phase velocities associated with these
plasmon branches have opposing signs at a common frequency.

\section{General Formulation of the Problem}
\label{sec2}

\begin{figure}[t]
\centering
\includegraphics[width=0.45\textwidth]{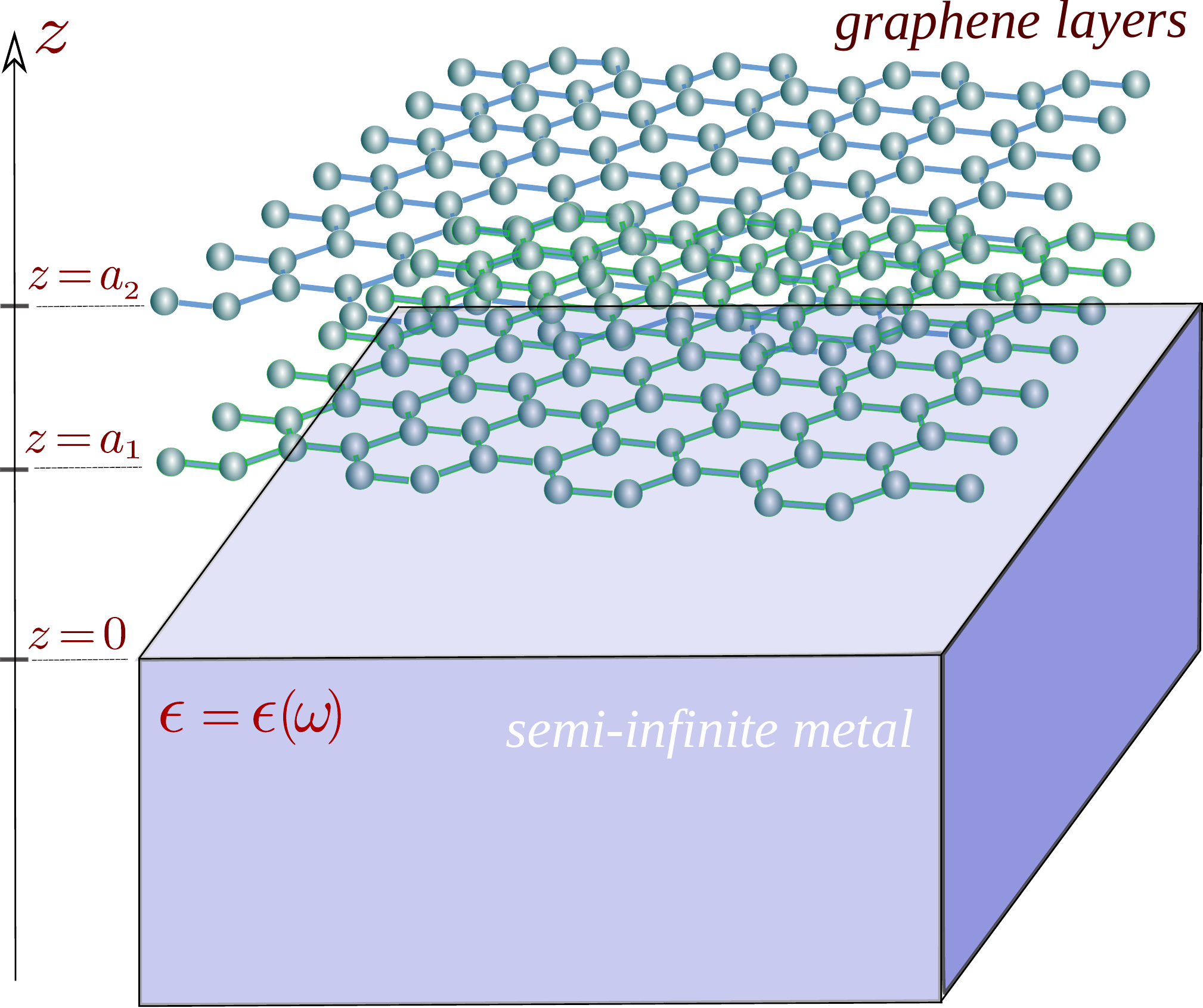}
\caption{(Color online)  Schematic illustration of a spiler generator consisting of a thick (semi-
infinite) conducting material on whose surface a plasmon resonance may induce  an instability by coupling
to the 2D plasmons on a pair of thin layers such as graphene, silicene or 2DEG at a hetero-interface.}
\label{FIG:1}
\end{figure}

In our formalism, we consider a nano-scale system consisting
of a pair of 2D layers and a thick conducting material.  The
layer may be monolayer graphene or a 2DEG such as a
semiconductor  inversion layer or HEMT (high electron mobility
transistor).  The graphene layer may have a gap, thereby
extending the flexibility of the composite system that
also incorporates a thick layer of conducting material as
depicted in Fig.\,\ref{FIG:1}. The excitation spectra
of allowable modes will be determined from a knowledge
of the non-local dielectric function
$\epsilon ({\bf r},{\bf r}^\prime;\omega)$ which depends
on position coordinates ${\bf r}, {\bf r}^\prime$ and
frequency $\omega$ or its inverse
$ K({\bf r},{\bf r}^\prime;\omega)$ satisfying
$\int d{\bf r}^\prime\  K({\bf r},{\bf r}^\prime;\omega)\,\epsilon({\bf r}^\prime,{\bf r}^{\prime\prime};\omega)
=\delta({\bf r}-{\bf r}^{\prime \prime})$.  The self-consistent
field equation for $K({\bf r},{\bf r}^\prime;\omega)$ is now determined.

In operator notation, the dielectric function  for the 2D layer and semi-infinite
structure is given by

\begin{equation}
\hat{\epsilon}= \hat{1}+\hat{\alpha}_{SI}+\hat{\alpha}_{2D} \equiv \hat{\epsilon}_{SI}+ \hat{\alpha}_{2D}
=\hat{K}_{SI}^{-1}+  \hat{\alpha}_{2D}\ ,
\end{equation}
where $\hat{\epsilon}=\hat{K}^{-1}$ with $\hat{K}$ as the inverse dielectric function satisfying

\begin{equation}
\hat{K}=\hat{K}_{SI} -\hat{K}_{SI} \cdot  \hat{\alpha}_{2D}\cdot  \hat{K}\ .
\end{equation}
Here, $\hat{\alpha}$, $\hat{\alpha}_{2D}$ and $\hat{\alpha}_{SI}$ are the polarization functions
of the composite system, the polarization function of the 2D layer
and semi-infinite substrate, respectively. Additionally, $\hat{K}_{SI}$ is the
inverse dielectric function for the semi-infinite substrate whose surface lies in the $z=0$ plane.
In integral form, after Fourier transforming with respect to coordinates
parallel to the $xy$-plane and
suppressing the in-plane wave number $q_{||}$ and frequency $\omega$,
we obtain

\begin{equation}
K(z_1,z_2)= K_{SI}(z_1,z_2) - \int_{-\infty}^\infty dz^\prime \int_{-\infty}^\infty
dz^{\prime\prime}\  K(z_1,z^\prime)\,
\alpha_{2D}(z^\prime ,z^{\prime\prime})\,K(z^{\prime\prime} ,z_2)\ .
\end{equation}
Here, the polarization function for the 2D structure is given by

\begin{equation}
\alpha_{2D}(z^{\prime}, z^{\prime\prime})= \int_{-\infty}^\infty  dz^{\prime\prime\prime} \
v(z^\prime-z^{\prime\prime\prime})\,D(z^{\prime\prime\prime},z^{\prime\prime})\ ,
\end{equation}
where $v(z-z^{\prime})=(2\pi e^2/\epsilon_s q_\parallel)\,\exp(-q_\parallel|z-z^\prime|)$, $\epsilon_s=4\pi \epsilon_0\epsilon_r$, and the 2D response function obeys

\begin{equation}
D(z^{\prime\prime\prime},z^{\prime\prime}) = \Pi_{2D}^{(0)} (q_{||},\omega)\,
\delta(z^{\prime\prime\prime}-a)\,\delta(z^{\prime\prime}-a)
\end{equation}
with $\Pi_{2D}^{(0)} (q_{||},\omega)$ as the single-particle in-plane response.  Upon
substituting this form of the polarization function for the monolayer into
the integral equation for the inverse dielectric function, we have

\begin{equation}
K(z_1,z_2)= K_{SI}(z_1,z_2) -
\Pi_{2D}^{(0)} (q_{||},\omega) \int_{-\infty}^\infty  dz^\prime\ K_{SI}(z_1,z^\prime)\,v(z^\prime-a)\,K(a,z_2)\ .
\label{eq:GG}
\end{equation}
We now set $z_1=a$ in Eq.\,(\ref{eq:GG}) and obtain

\begin{equation}
K(a,z_2)= K_{SI}(a,z_2) -
\Pi_{2D}^{(0)} (q_{||},\omega) \left\{\int_{-\infty}^\infty  dz^\prime\ K_{SI}(a,z^\prime)\,v(z^\prime-a)\right\}
K(a,z_2)\ .
\label{eq:GG2}
\end{equation}
Solving for $K(a,z_2)$ yields

\begin{equation}
K(a,z_2) = \frac{ K_{SI}(a,z_2)}{S_{C}(q_{||},\omega)}\ ,
\label{eq:GG3}
\end{equation}
and

\begin{equation}
S_{C}(q_{||},\omega) \equiv 1+ \Pi_{2D}^{(0)} (q_{||},\omega) \left\{\int_{-\infty}^\infty  dz^\prime\
 K_{SI}(a,z^\prime)\,v(z^\prime-a)\right\}
\end{equation}
whose zeros determine the plasmon resonances. In our numerical calculations, we shall use
$K_{SI} (z,z^\prime)$ given in Eq.\,(30) of Ref.\,[\onlinecite{Horing}].
Thus, from Eq.\,(\ref{eq:GG}), we obtain

\begin{equation}
K(z_1,z_2)= K_{SI}(z_1,z_2) -
\Pi_{2D}^{(0)} (q_{||},\omega)\frac{ K_{SI}(a,z_2)}{S_{C}(q_{||},\omega)}
\left\{ \int_{-\infty}^\infty \  dz^\prime\ K_{SI}(z_1,z^\prime) v(z^\prime-a) \right\}  \
\ .
\label{eq:GG-inv}
\end{equation}
\medskip

In the local limit, we have

\begin{equation}
S_{C}(q_{||},\omega)= 1+\frac{2\pi e^2}{\epsilon_s q_{||}}\,\Pi_{2D}^{(0)} (q_{||},\omega)
\left\{ 1+ e^{-2q_{||} a}  \frac{1-\epsilon_B(\omega)}{1+\epsilon_B(\omega)}   \right\}\ .
\end{equation}

With $\Pi_{2D}^{(0)} (q_{||},\omega)\approx Cq_\parallel^2/\omega^2$, we obtain
the following equation:
\begin{equation}
1-\frac{2 \pi  C e^2}{\epsilon_s \omega^2}q_{\parallel} \left\{
1+  e^{-2q_{\parallel}a}\,\frac{\omega_p^2}{2 \omega^2 - \omega_p^2}
\right\} = 0 \, ,
\end{equation}
which is a quadratic equation for $\omega^2$. For a 2DEG, we have $C=n_{2D}/m_{2D}^\ast$.
For graphene,

\begin{equation}
C= \frac{2\mu}{\pi\hbar^2}\left\{ 1-\frac{\Delta^2}{\mu^2} \right\}\ ,
\end{equation}
where $\mu$ is the chemical potential and $\Delta$ is the gap
between valence and conduction bands.
Consequently, we find the plasmon frequency as follows:\,\cite{NJMH}

\begin{equation}
\omega^2= K_1 \pm \sqrt{K_2}
\end{equation}
with $K_1$ and $K_2$ defined by:

\begin{eqnarray}
&&  K_1 = \frac{\pi e^2  C  }{\epsilon_s} q_{\parallel}
+\left(\frac{\omega_p}{2} \right)^2\ ,
\nonumber\\
&& K_2 = \frac{\pi e^2  C \, \omega_p^2}{\epsilon_s}
 e^{-2q_{\parallel}a}\,q_{\parallel} +\left[
\left(\frac{\omega_p}{2} \right)^2 - \frac{C e^2 \pi }{\epsilon_s} q_{\parallel} \right]^2 \, .
\end{eqnarray}
In the long-wavelength limit $(q_\parallel \ll k_F)$ these expressions are reduced to:

\begin{eqnarray}
&& \omega_1^2 \approx \frac{4 \pi  C e^2 a}{\epsilon_s} q_{\parallel}^2\ ,  \\
\nonumber
&& \omega_2^2 \approx  \frac{\omega_p^2}{2} + \frac{2 \pi  C e^2 }{\epsilon_s} q_{\parallel}\ ,
\end{eqnarray}
and the frequencies are

\begin{eqnarray}
&& \omega_1 \approx 2 e \sqrt{\frac{\pi a  C }{\epsilon_s}}\,q_{\parallel}\ ,   \\
\nonumber
&& \omega_2 \approx \frac{\omega_p}{\sqrt{2}} + \frac{\sqrt{2} \pi  C e^2 }{\epsilon_s \omega_p}\,q_{\parallel}\ ,
\end{eqnarray}
which are both linear in $q_{||}$, unlike the $q_{||}^{1/2}$-dependence for free-standing
graphene or the 2DEG\,\cite{7,8,9,10,wunch,pavlo}.
\medskip

We may generalize the formalism to a structure with a double layer positioned
at $z=a_1$ and $z=a_2$  ($0<a_1<a_2$)  interacting with each other as  well as the semi-infinite
conducting substrate with its surface located at $z=0$. A similar calculation shows that

\begin{eqnarray}
K(z_1,z_2) &=& K_{SI}(z_1,z_2) -
\Pi_{2D;1}^{(0)} (q_{||},\omega) \int_{-\infty}^\infty
 dz^\prime\ K_{SI}(z_1,z^\prime) v(z^\prime-a_1) K(a_1,z_2)
\nonumber\\
&-&    \Pi_{2D;2}^{(0)} (q_{||},\omega) \int_{-\infty}^\infty
 dz^\prime\ K_{SI}(z_1,z^\prime) v(z^\prime-a_2) K(a_2,z_2)\ .
\label{eq:GGDH}
\end{eqnarray}
By setting $z_1=a_1$ and $z_1=a_2$ in turn in Eq.\,(\ref{eq:GGDH}) and solving the
pair of simultaneous equations for $K(a_1,z_2)$ and $K(a_2,z_2)$, we obtain

\begin{equation}
\left[
\begin{matrix} K(a_1,z_2)\cr
K(a_2,z_2)\cr
\end{matrix}
\right]= \frac{1}{S_c^{(2)}(q_{||},\omega)} \tensor{{\cal M}} (q_{||},\omega)
\left[
\begin{matrix} K_{SI}(a_1,z_2)\cr
K_{SI}(a_2,z_2)\cr
\end{matrix}
\right]\ ,
\end{equation}
where $S_c^{(2)}(q_{||},\omega)=\mbox{Det} \tensor{{\cal M}} (q_{||},\omega)$ with

\[
\tensor{{\cal M}}(q_{||},\omega)=
\]
\begin{equation}
\left[ \begin{array}{cc}
1+\Pi_{2D;2}^{(0)}(q_{||},\omega)\int\limits_{-\infty}^\infty dz^\prime K_{SI}(a_2,z^\prime)v(z^\prime-a_2) &
-\Pi_{2D;2}^{(0)}(q_{||},\omega)\int\limits_{-\infty}^\infty dz^\prime K_{SI}(a_1,z^\prime)v(z^\prime-a_2) \cr
-\Pi_{2D;1}^{(0)}(q_{||},\omega)\int\limits_{-\infty}^\infty dz^\prime K_{SI}(a_2,z^\prime)v(z^\prime-a_1) &
1+\Pi_{2D;1}^{(0)}(q_{||},\omega)\int\limits_{-\infty}^\infty dz^\prime K_{SI}(a_1,z^\prime)v(z^\prime-a_1)
\end{array}
\right]\ .
\end{equation}
Substituting the results for $K(a_1,z_2)$ and $K(a_2,z_2)$ into Eq.\,(\ref{eq:GGDH}),
we obtain the complete inverse dielectric function for a pair
of 2D planes interacting with each other and a semi-infinite conducting material.
The plasmon excitation frequencies are determined by the zeros of
$S_c^{(2)}(q_{||},\omega)$. Furthermore, the effect of the inverse dielectric
function for the semi-infinite structure  $K_{SI} (z,z^\prime;q_{||},\omega)$ screens
coupling within and between two layers. As a matter of fact, our result
for the plasmon dispersion relation generalizes that obtained by Das Sarma and
Madhukar\,\cite{DasSarma} for a bi-plane. In the current case, we obtain in the local limit

\begin{eqnarray}
&&S_c^{(2)}(q_{||},\omega) =\left\{  1+  \frac{2\pi e^2}{\epsilon_s q_{||}}\,\Pi_{2D;2}^{(0)} (q_{||},\omega)
\left [  1  +  e^{-2q_{||} a_2 }\,\frac{\omega_p^2}{2\omega^2-\omega_p^2} \right]
\right\}
\nonumber\\
&\times&
\left\{  1+  \frac{2\pi e^2}{\epsilon_s q_{||}}\,\Pi_{2D;1}^{(0)} (q_{||},\omega)
\left [  1  +  e^{-2 q_{||} a_1 }\,\frac{\omega_p^2}{2\omega^2-\omega_p^2} \right]
\right\}
\nonumber\\
&-&   \left(  \frac{2\pi e^2}{\epsilon_s q_{||}} \right)^2
\Pi_{2D;1}^{(0)} (q_{||},\omega)\,\Pi_{2D;2}^{(0)} (q_{||},\omega)
\left [  e^{-q_{||} |a_1-a_2|}   +  e^{- q_{||} (a_1 +a_2) }\,\frac{\omega_p^2}{2\omega^2-\omega_p^2}
\right]^2\ .
\end{eqnarray}
\medskip

We now introduce our notation, $\overline{C}_j=2\pi e^2C_j/(\epsilon_s\omega_p^2)$ for $j=1,2$.
The spectral function yields  real frequencies. A plane interacting with
the half-space has two resonant modes. Each pair of 2D layers interacting in isolation
far from the half-space conducting medium supports a symmetric and an
antisymmetric mode\,\cite{DasSarma}.  In the absence of a driving current, the analytic solutions  for the plasmon modes for a
pair of 2D layers that are Coulomb coupled to a half-space are given by

\begin{eqnarray}
&& \Omega_{1}(q_\parallel)/\omega_p = 1/\sqrt{2} + q_\parallel\,(\overline{C}_1+\overline{C}_2)/\sqrt{2} + \mathcal{O}[q_\parallel^2]\ , \\
\nonumber
&& \Omega_{2}(q_\parallel)/\omega_p =  q_\parallel \sqrt{\overline{C}_1 a_1 + \overline{C}_2 a_2 + \sqrt{\mathcal{A} }}+ \mathcal{O}[q_\parallel^2]\ , \\
\nonumber
&& \Omega_{3}(q_\parallel)/\omega_p =  q_\parallel \sqrt{\overline{C}_1 a_1 + \overline{C}_2 a_2 - \sqrt{\mathcal{A} }} + \mathcal{O}[q_\parallel^2]\ ,
\label{solutions}
\end{eqnarray}
where $\mathcal{A} \equiv (\overline{C}_1 a_1 -\overline{C}_2 a_2)^2 + 4 \overline{C}_1 \overline{C}_2 a_1^2$ 
and the term $4 \overline{C}_1 \overline{C}_2 a_1^2$ plays the role of ``Rabi coupling''.
Clearly, for long wavelengths, only $\Omega_1(q_\parallel)$ depends on $\omega_p$.\
However, the excitation spectrum changes dramatically when a current is driven
through the configuration.  Under a constant electric field, the carrier distribution
is modified, as may be obtained by employing the relaxation time approximation
in the equation of motion for the center-of-mass momentum. For carriers in a parabolic energy band
with effective mass $m^\ast$ and  drift velocity ${\bf v}_d$ determined by
the electron mobility and the external electric field, the electrons in the medium
are redistributed. This is determined by  a momentum shift in the
wave vector ${\bf k}_\parallel \to {\bf k}_\parallel -m^\ast {\bf v}_d/\hbar$
in the thermal-equilibrium  energy
distribution function $f_0(\epsilon_{\bf k} )$. By making a change of variables
in the well-known Lindhard polarization function $\Pi^{(0)}_{2D}(q_\parallel,\omega)$, this effect is
equivalent to a frequency shift $\omega\to \omega-{\bf q}_\parallel\cdot {\bf v}_d$.
For massless Dirac fermions in graphene with linear energy dispersion, this Doppler
shift in frequency is not in general valid for arbitrary wave vector as we prove
in our Appendix. This is our conclusion after we relate
the surface    current density to the center-of-mass wave vector in a steady state.
Our calculation shows that the redistribution of electrons  leads to a shift in the
wave vector appearing in the Fermi  function by the center-of-mass wave vector
${\bf K}_0=(k_F/v_F){\bf v}_d$ where $k_F$ and $v_F$ are the Fermi wave vector and
velocity, respectively.  However, in the long wavelength limit, $q_\parallel\to 0$, the Doppler
shift in frequency is approximately obeyed. This is discussed in Appendix A.
 Consequently, regardless of the nature
of the 2D layer represented in the dispersion equation we may replace
$\omega\to \omega-{\bf q}_\parallel\cdot {\bf v}_d$ in the dispersion equation in the presence
of an applied  electric field at long wavelengths.
\medskip

We shall treat the solution frequencies $\omega_{\pm}(q_\parallel)$ as complex variables with
${\rm Im}[\omega_{\pm}(q_\parallel)]\geq 0$, where ${\rm Im}[\omega_{\pm}(q_\parallel)]>0$ implies a
finite growth rates $\gamma_{\pm}(q_\parallel)={\rm Im}[\omega_{\pm}(q_\parallel)]$ for two split plasmon modes.
Since $\epsilon(q_\parallel,\,\omega)$ is a complex function, we ask for
${\rm Re}[\epsilon(q_\parallel,\,\omega)]={\rm Im}[\epsilon(q_\parallel,\,\omega)]=0$.
Therefore, we are left with damping-free plasmon modes in the system but
they still face possible instability due to ${\rm Im}[\omega_{\pm}(q_\parallel)]>0$.

\section{Numerical  Results and Discussion}
\label{sec3}

\begin{figure}[t]
\centering
\includegraphics[width=0.6\textwidth]{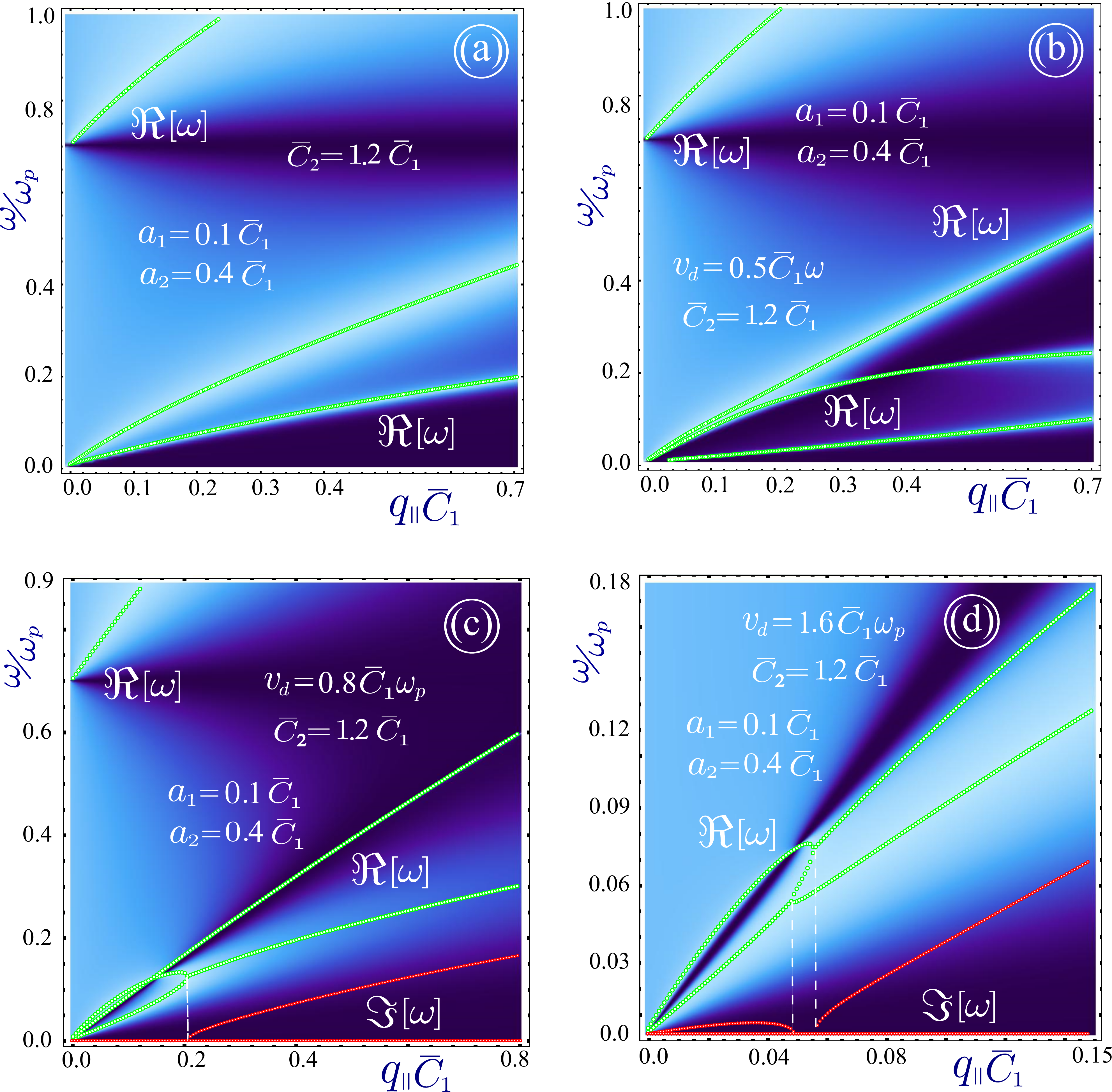}
\caption{(Color online) Complex frequencies $\omega$ yielding the plasmon dispersion
(real part $\mathfrak{R}[\omega]$) and inverse  lifetime (imaginary part $\mathfrak{I}[\omega]$)
for a pair of 2D layers and a semi-infinite conducting medium as presented in
Fig.\,\ref{FIG:1}. The plasma frequency for the semi-infinite medium is $\omega_p$.
The layers are located at $a_1=0.1\,\bar{C}_1$ and $a_2=0.4\,\bar{C}_1$
with respect to the surface. Each panel corresponds to a different drift velocity
$v_d/\omega_p=0$, $0.5\,\bar{C}_1$, $0.8\,\bar{C}_1$ and $1.6\,\bar{C}_1$.
Panel $(a)$ with $v_d=0$ corresponds to the solutions in Eq.\,(\ref{solutions}).
Panel $(b)$ shows that for a small drift $v_d$  additional plasmon branch appears, but all  solutions are stable.
The carrier concentrations,  chemical potentials or temperatures in the layers  are such that
$\bar{C}_1 = 1.2 \, \bar{C}_2$ for all cases. Either of the two lowest
plasmon branches might become unstable, depending on $v_d$. The Rabi splitting
of the plasmon excitation branches  by the external electric field  is attributed to
quasiparticles with different excitation energies for the same wavelength. After the ``loop" closes
in $(c)$, the lowest branch becomes unstable with finite imaginary part illustrated by
the red curve.  In $(d)$, the lowest plasmon branch has an instability for two
separate ranges of wave vector.}
\label{FIG:2}
\end{figure}

First, we have investigated numerically the effect of passing a current through a layer of
2DEG, graphene  or silicene in the presence of a surface, for a pair of 2D layers
and a semi-infinite conducting medium as presented in Fig.\,\ref{FIG:1}.
In Fig.\,\ref{FIG:2}, we present the
complex frequencies which yield the plasmon dispersion
(real part) and inverse  lifetime (imaginary part).
The layers are located at $z=a_1 = 0.1 \, \bar{C}_1$ and $z=a_2 = 0.4 \, \bar{C}_1$.
 Each panel presents results for a different drift velocity
$v_d/\omega_p = 0$, $0.5\,\bar{C}_1$, $0.8\,\bar{C}_1$
and $1.6\,\bar{C}_1$. In the absence of a drift current, ($v_d=0$), panel $(a)$ shows that there are three
plasmon branches excited, which are stable as given by Eq.\,(\ref{solutions}). At low drift velocity, panel $(b)$ shows that the
plasmons are still stable. However, as the current is increased, the lowest branches may become unstable as demonstrated in panels $(c)$ and $(d)$ through the appearance of a finite imaginary part for the frequency.
The carrier concentrations, chemical potentials or temperatures in the layers are such that
$\bar{C}_1 = 1.2 \, \bar{C}_2$ for all cases. Either of the two lowest
plasmon branches might become unstable, depending on the strength of the
drift current. The Rabi splitting of the plasmon  branches in the presence of
an external electric field  is a consequence of quasiparticles having different
energies for the same wavelength.
\medskip

From the perspective of real space, the plasmon wave is a longitudinal charge density
wave with both its density fluctuation and phase velocity $v_p$ along the ${\bf q}_\parallel$ (propagation) direction.
When there are two electron gas
layers separated by a distance $\Delta a$ in the $z$ direction, the inter-layer coulomb
coupling will produce two plasmon modes with different energy dispersions
$\omega_{\pm}(q_\parallel)$ and phase velocities
$v_{\pm}(q_\parallel)=\omega_{\pm}(q_\parallel)/q_\parallel$,
which are separately associated with the symmetric [in-phase, upper $\omega_+(q_\parallel)$ branch in Fig.\,\ref{FIG:2}(a)] and
antisymmetric  [out-of-phase, lower $\omega_-(q_\parallel)$ branch] states of charge-density waves.
The symmetric plasmon mode gives rise to a dipole-like plasmon excitation, similar
in nature to that of a single layer with an electron density given by adding the
densities on the two layers.  On the other hand, the antisymmetric plasmon mode
leads to a quadruple-like excitation, which is able to store electric field energy
between the two layers. In the presence of a current with drift velocity $v_d$ for
electrons in one  of the two electron-gas layers, the phase velocities of these two
spatially separated  plasmon modes are modified drastically
due to the Doppler shift as well as the surface plasmon on the substrate.
Consequently,  two streams of quasiparticles with different phase velocities may
be created from either the symmetric or antisymmetric mode, depending on the strength
of the current, the separation between layers and their distances from the surface of the substrate.
However, for freely suspended layers,  the lower-energy antisymmetric plasmon mode
becomes unstable whenever $v_d$ lies within  the range   $v_{-}(q_\parallel)<v_d<v_{+}(q_\parallel)$, 
switching from a quadruple-like excitation to a dipole-like
excitation. The instability for the   plasmon mode has a finite lifetime  because it
grows after taking energy from the injected current and  transferring it to the incident
electromagnetic field.
\medskip

From the point of view of momentum space, electrons may only occupy   momentum
space within the range of $|{\bf k}_\parallel|\leq k_F$ at zero temperature in a state of thermal
equilibrium,   where $k_F$ is the
electron Fermi wave number and $\varepsilon(k_F)=\varepsilon(-k_F)=E_F$ is the Fermi
energy.  When a current is passed through the electron gas, electrons are driven out
from this state of thermal equilibrium  and their population becomes asymmetrical
with respect to $k_\parallel=0$.    In this case, the Fermi energy $E_F$ is split into $E_{F,+}=\varepsilon(k_F+K_c)$ and $E_{F,-}=\varepsilon(-k_F+K_c)$ with $E_{F,+}>E_{F,-}$, where $\hbar K_c$ represents
the electron center-of-mass momentum. In this shifted Fermi-Dirac distribution model,
electrons in such a non-equilibrium state are energetically unstable, and the higher-energy
electrons in the range   $k_F\leq k_\parallel\leq k_F+K_c$ tend to relax to the
lower-energy empty states in the range of $-k_F\leq k_\parallel\leq K_c-k_F$ by emitting
electromagnetic waves and phonons    to ensure   total momentum and energy conservation.
This process is known as   radiation loss of plasmon excitations in the time domain, in addition to
the usual absorption loss of plasmon excitations in the space domain due to nonzero
imaginary part of the electron dielectric function.
\medskip

A surface grating must be employed in order to convert the energy of the
unstable  plasmon mode into a transverse radiation field in
free space and at the same time  suppress   phonon emission (heating).
The growth rate $\gamma_{p}(q_\parallel)={\rm Im}[\omega_{p}(q_\parallel)]$ of the
plasmon mode is  determined by the imaginary part of the plasmon frequency
while the average plasmon growth rate (per unit area) is given by
$\gamma(K_c)=(2\pi)^{-1}\int\limits_0^{\infty}\,dq_\parallel\ \gamma_{p}(q_\parallel)\,q_\parallel$.
\medskip

\section{Concluding Remarks}
\label{sec4}

In summary, we are proposing a spiler  quantum plasmonic device  which employs
2D layers in combination with a thick conducting material. We find that the spiler emits
electromagnetic radiation when a current is  passed through  the 2D layer or the surface
of the thick conducting material  to make the plasmons become unstable at a
specific frequency and wave number.  It is possible to change the range of plasmon
instability by selecting the properties of the nanosheet or frequency of the
surface plasmon, i.e., the substrate.  The surface plasmon plays a crucial role
in giving rise to the Rabi splitting and the concomitant streams of quasiparticles
whose phase velocities are in opposite directions when the instability takes place.
The emitted electromagnetic radiation may be collected from regions on the surface that are
convenient. Finally, we note that in  presenting our numerical results, we  measured
frequency  in terms of the bulk plasmon frequency which, typically for heavily-doped semiconductors, we have
$\hbar\omega_p\sim  0.5$\,eV. Either for intrinsic graphene, doped monolayer graphene
or an inversion layer 2DEG,  we have $\overline{C}\sim 10^{-4}$\,m and $v_d\sim 10^6$\,m/s.
In Fig.\,\ref{FIG:2}, the unit of frequency is $\omega_0=\sqrt{2\pi e^2 C_1 k_F/\epsilon_s}$
which is of the same order as $\omega_p$.
\medskip

The current-driven asymmetric electron distribution in $k$ space leads to an induced
polarization current or a ``dipole radiator''. If two electron gas layers are
placed  close enough, the in-phase inter-layer Coulomb interaction will give a
dipole-like plasmon excitation, similar to that of a single layer. On the other hand,
the out-of-phase  layer Coulomb coupling  will lead to an unstable quadruple-like excitation.
This unstable quadruple-like plasmon excitation can be effectively converted into a transverse
electromagnetic field in   free space if a surface grating is employed.

\acknowledgments
This research was supported by  contract \# FA 9453-13-1-0291 of
AFRL.  DH would like
to thank the Air Force Office of Scientific Research (AFOSR) for its support.
We acknowledge having helpful discussions with  Oleksiy Roslyak and Antonios Balassis.

\appendix

\section{Effect of Drift Current on the Polarization}

Let us first consider the case when a current is passed through a 2DEG layer.  This current
creates a Doppler shift in the response function $\Pi(q,\omega-\textbf{q}\cdot \textbf{v}_d)$.
The derivation is presented in the paper by Kempa, et al.\,\cite{Bakshi} and the argument is as
follows. The energy dispersion for an electron in a 2DEG is $\varepsilon_\textbf{k}=\hbar^2\textbf{k}^2/2m^\ast$,
where $m^\ast$ is the electron effective  mass. The current flow leads to the replacement in wave vector
$\textbf{k}\to \textbf{k}-m^\ast \textbf{v}_d/\hbar$ everywhere in the polarization function

\begin{equation}
\Pi({\bf q},\,\omega)=2\,\int\,\frac{d^2{\bf k}}{(2\pi)^2}\,
\, \frac{f_0(\varepsilon_{|{\bf k}-{\bf q}|},\,T)-f_0(\varepsilon_k,\,T)}
{\varepsilon_{|{\bf k}|}-\varepsilon_{|{\bf k}-{\bf q}|}-\hbar(\omega+i0^+)} \ ,
\label{kempa}
\end{equation}
where $f_0(\varepsilon)$ is the equilibrium distribution function for electrons. After this
wave vector replacement is carried out and a change of variables is made in the
resulting integral, we simply obtain  an expression for the polarization function which
is exactly the same as that given in Eq.\,(\ref{kempa}), except with the frequency
shifted by $\textbf{q}\cdot \textbf{v}_d$.
\medskip

We now turn to the case of graphene which is characterized by massless Dirac fermions
for which  the energy dispersion is linear in the wave vector $\textbf{k}$.
For a spatially-uniform system, the first-order moment of the Boltzmann equation for a
single valley gives us

\begin{equation}
\frac{\partial{\bf j}(t)}{\partial t}=-\frac{{\bf j}(t)}{\tau_0}
-N_c\,\frac{\nu_0^2e}{2}\,{\bf F}(t)\,(k_BT)\,{\cal Q}_0(\eta)\ ,
\end{equation}
where $N_c=1/(\pi\hbar^2\nu_0^2)$, $\eta=\mu_0(t)/k_BT$, $\mu_0$ is the chemical
potential of electrons in graphene, the quantity

\begin{equation}
{\cal Q}_0(\eta)=\int\limits_0^\infty\,\frac{dx}{e^{(x-\eta)}+1}\ ,
\end{equation}
and

\begin{equation}
{\bf j}(t)=\frac{2}{{\cal A}}\,\sum\limits_{{\bf k}}\,{\bf v}_k\,f_0(\varepsilon_k,\,T,\,t)\ ,
\end{equation}
is the electron surface current density,  ${\cal A}$ is the sample area,
$T$ is the electron temperature, $\hbar{\bf k}$ is the electron wave vector,
$\varepsilon_k=\hbar\nu_0\,k$ is the electron kinetic energy,
$\nu_0$ is the Fermi velocity of graphene, ${\bf v}_k=\nabla_k\varepsilon_k/\hbar$
is the electron group velocity, and $\tau_0$ is the average momentum-relaxation
time, ${\bf F}(t)$ is the external electric field. Additionally,
we have the following relation

\begin{equation}
-N_c\,(k_BT)\,{\cal Q}_0(\eta)=\frac{2}{{\cal A}}\,\sum\limits_{{\bf k}}\,
\frac{\partial f_0(\varepsilon_k,\,T,\,t)}{\partial\varepsilon_k}\ ,
\end{equation}

\begin{equation}
\rho(t)=\frac{2}{{\cal A}}\,\sum\limits_{{\bf k}}\,f_0(\varepsilon_k,\,T,\,t)\ ,
\end{equation}
where $\rho(t)$ is the electron areal density. Considering a steady state under
a constant electric field ${\bf F}_0$, we obtain

\begin{equation}
{\bf j}_0=\frac{\hbar\nu_0^2}{2}\,\frac{e\tau_0}{\hbar}\,{\bf F}_0\,
\frac{2}{{\cal A}}\,\sum\limits_{{\bf k}}\,
\frac{\partial f_0(\varepsilon_k,\,T)}{\partial\varepsilon_k}\ .
\end{equation}
At $T\approx 0$\,K, we have

\begin{equation}
\frac{2}{{\cal A}}\,\sum\limits_{{\bf k}}\,
\frac{\partial f_0(\varepsilon_k,\,T)}{\partial\varepsilon_k}
\approx-\frac{E_F}{\pi\hbar^2\nu_0^2}\ ,
\end{equation}
where $E_F=\hbar\nu_0\,k_F$ is the electron Fermi energy and $k_F=\sqrt{2\pi\rho_0}$
is the Fermi wave number. As a result, this leads to

\begin{equation}
{\bf j}_0=\frac{E_F}{2\pi\hbar}\,\frac{e\tau_0}{\hbar}\,{\bf F}_0\ .
\end{equation}
\medskip

In the second-quantization picture, the Hamiltonian operator for $N$ electrons in
graphene in the presence of an  field may be written as

\begin{equation}
\hat{{\cal H}}(t)=\sum\limits_{j=1}^N\,\nu_0\,\vec{\sigma}
\cdot\hat{\bf p}_j-\sum\limits_{j=1}^N\,e\,{\bf F}(t)\cdot{\bf r}_j\ ,
\end{equation}
where $\vec{\sigma}=(\sigma_x,\,\sigma_y)$ is the Pauli-matrix vector, $\hat{\bf p}_j=-i\hbar\nabla_j$ is
the electron momentum operator, and ${\bf r}_j$ is the electron position vector.
For this system, we   define the center-of-mass momentum operator as
$\hat{\bf P}_c=\sum\limits_{j=1}^N\,\hat{\bf p}_j$. Therefore, the Heisenberg equation gives us

\begin{equation}
\frac{d\hat{\bf P}_c(t)}{dt}=-\frac{\hat{\bf P}_c(t)}{\tau_0}+\frac{1}{i\hbar}\,\left[\hat{\bf P}_c(t),\,\hat{{\cal H}}(t)\right]=-\frac{\hat{\bf P}_c(t)}{\tau_0}+\frac{1}{i\hbar}\,\sum\limits_{j=1}^N\,\left[\hat{\bf p}_j,\,\hat{{\cal H}}(t)\right]
=-\frac{\hat{\bf P}_c(t)}{\tau_0}+e{\bf F}(t)\ ,
\end{equation}
where we have employed the momentum-relaxation time approximation. For a steady
state, we have

\begin{equation}
{\bf K}_0\equiv\frac{{\bf P}_0}{\hbar}=\frac{e\tau_0}{\hbar}\,{\bf F}_0\ .
\end{equation}
Finally, we are able to connect the electron surface current density with  the
center-of-mass wave vector in a steady state simply through

\begin{equation}
{\bf j}_0=\frac{E_F}{2\pi\hbar}\,{\bf K}_0\ .
\end{equation}
Recalling that we have ${\bf j}=\rho_0\,{\bf v}_d$, where ${\bf v}_d$ is the
drift velocity of electrons in the system, we arrive at the relation

\begin{equation}
{\bf K}_0=\frac{2\pi\hbar\rho_0}{E_F}\,{\bf v}_d
=\frac{\hbar k_F^2}{E_F}\,{\bf v}_d=\frac{k_F}{\nu_0}\,{\bf v}_d\ .
\end{equation}
Consequently, for   drifted electrons we find from the Lindhardt  polarization
function that

\[
\Pi({\bf q},\,\omega)=2\,\int\,\frac{d^2{\bf k}}{(2\pi)^2}\,
\left\{1+\cos[\Theta_{{\bf k}{\bf q}}({\bf v}_d)]\right\}\,
\frac{f_0(\varepsilon_{|{\bf k}-{\bf q}|},\,T)-f_0(\varepsilon_k,\,T)}
{\varepsilon_{|{\bf k}+{\bf K}_0|}-\varepsilon_{|{\bf k}-{\bf q}+{\bf K}_0|}-\hbar(\omega+i0^+)}
\]
\begin{equation}
=2\,\int\,\frac{d^2{\bf k}}{(2\pi)^2}\,\left\{1+\cos[\Theta_{{\bf k}{\bf q}}({\bf v}_d)]\right\}\,
\frac{f_0(\varepsilon_{|{\bf k}-{\bf q}|},\,T)-f_0(\varepsilon_k,\,T)}
{\hbar\nu_0k_F[{\cal S}({\bf k},\,{\bf v}_d)-{\cal S}({\bf k}-{\bf q},\,
{\bf v}_d)]-\hbar(\omega+i0^+)}\ ,
\end{equation}
where ${\bf v_d}$ is determined by the product of the electron mobility and the
external electric field ${\bf F}_0$, and

\begin{equation}
{\cal S}({\bf k},\,{\bf v}_d)=\sqrt{(k/k_F)^2+(v_d/\nu_0)^2+2({\bf k}/k_F)
\cdot({\bf v}_d/\nu_0)}\ ,
\end{equation}

\begin{equation}
\cos[\Theta_{{\bf k}{\bf q}}({\bf v}_d)]=\frac{[({\bf k}/k_F)+({\bf v}_d/
\nu_0)]\cdot[({\bf k}/k_F)-({\bf q}/k_F)+({\bf v}_d/\nu_0)]}
{{\cal S}({\bf k},\,{\bf v}_d)\,{\cal S}({\bf k}-{\bf q},\,{\bf v}_d)}\ .
\end{equation}
When $v_d/\nu_0\ll 1$ is satisfied, we obtain

\begin{equation}
{\cal S}({\bf k},\,{\bf v}_d)\approx\frac{k}{k_F}+\left(\frac{{\bf k}}{k}\right)
\cdot\left(\frac{{\bf v}_d}{\nu_0}\right)+{\cal O}
\left[\left(\frac{v_d}{\nu_0}\right)^2\right]\ ,
\end{equation}

\begin{equation}
{\cal S}({\bf k},\,{\bf v}_d)-{\cal S}({\bf k}-{\bf q},\,{\bf v}_d)
\approx\left(\frac{{\bf q}}{k}\right)\cdot\left(\frac{{\bf v}_d}{\nu_0}\right)
+{\cal O}\left[\left(\frac{v_d}{\nu_0}\right)^2\right]\ ,
\end{equation}
and

\[
\cos[\Theta_{{\bf k}{\bf q}}({\bf v}_d)]\approx
\frac{\left[{\bf k}\cdot({\bf k}-{\bf q})+k_F(2{\bf k}-{\bf q})
\cdot({\bf v}_d/\nu_0)\right]k|{\bf k}-{\bf q}|}{k^2|{\bf k}-
{\bf q}|^2+k_F({\bf v}_d/\nu_0)\cdot
\left[k^2({\bf k}-{\bf q})+{\bf k}|{\bf k}-{\bf q}|^2\right]}
\approx\frac{{\bf k}\cdot({\bf k}-{\bf q})}{k|{\bf k}-{\bf q}|}
\]
\begin{equation}
+k_F\left(\frac{{\bf v}_d}{\nu_0}\right)\cdot\left\{\frac{2{\bf k}-{\bf q}}{k|{\bf k}-{\bf q}|}
-\left[\frac{{\bf k}\cdot({\bf k}-{\bf q})}{k|{\bf k}-{\bf q}|}\right]\left(\frac{{\bf k}-{\bf q}}{|{\bf k}-{\bf q}|^2}+\frac{{\bf k}}{k^2}\right)\right\}
+{\cal O}\left[\left(\frac{v_d}{\nu_0}\right)^2\right]\ .
\end{equation}

\end{document}